\documentclass[10pt,a4paper,twoside]{article}
\usepackage{epsfig}
\usepackage{baltlat6}
\usepackage{array}
\usepackage{here}
\begin{document}
\ \ \vspace{0.5mm} \setcounter{page}{305}

\titlehead{Baltic Astronomy, vol. 24, 305--313, 2015}

\titleb{STRUCTURAL PARAMETERS OF STAR CLUSTERS: \\ SIGNAL TO NOISE EFFECTS}

\begin{authorl}
\authorb{D. Narbutis}{},
\authorb{A. Brid{\v z}ius}{} and
\authorb{D. Semionov}{}
\end{authorl}

\begin{addressl}
\addressb{}{Center for Physical Sciences and Technology, \\ Savanori{\c u}
231, Vilnius LT-02300, Lithuania; donatas.narbutis@ftmc.lt}
\end{addressl}

\submitb{Received: 2015 September 28; accepted: 2015 October 15}

\begin{summary}
We study the impact of photometric signal to noise on the accuracy
of derived structural parameters of unresolved star clusters using
MCMC model fitting techniques. Star cluster images were simulated as
a smooth surface brightness distribution following a King profile
convolved with a point spread function. The simulation grid was
constructed by varying the levels of sky background and adjusting
the cluster's flux to a specified signal to noise. Poisson noise was
introduced to a set of cluster images with the same input parameters
at each node of the grid. Model fitting was performed using ``{\it
emcee}'' algorithm. The presented posterior distributions of the
parameters illustrate their uncertainty and degeneracies as a
function of signal to noise. By defining the photometric aperture
containing 80\% of the cluster's flux, we find that in all realistic
sky background level conditions a signal to noise ratio of $\sim$50
is necessary to constrain the cluster's half-light radius to an
accuracy better than $\sim$20\%. The presented technique can be
applied to synthetic images simulating various observations of
extragalactic star clusters.
\end{summary}


\resthead{Structural parameters of star clusters}{D. Narbutis, A.
Brid{\v z}ius, D. Semionov}

\sectionb{1}{INTRODUCTION}

We consider the simplest case of an unresolved star cluster detected
with an ideal CCD in a single image and located in a completely
empty environment. The only sources of photons are the cluster and
uniform sky background. Let the point spread function (PSF) be
known. We construct the cluster's model at position $(x, y)$ in the
image by using a King (1962) surface brightness profile defined by a
core radius, $r_{\rm c}$, and a tidal radius, $r_{\rm t}$. We
convolve it with the PSF, multiply by the cluster's flux, $f_{\rm
cl}$, and add sky background, $\mu_{\rm sky}$. This model image is
compared to the observed image to evaluate the goodness of the
model. The sky background usually is determined from pixels
surrounding the cluster and subtracted before model fitting.
However, we treat $\mu_{\rm sky}$ as free and perform complete
analysis of the recovered parameters.

The questions we address in this paper are: 1) what is the accuracy
of derived parameters $(x, y, r_{\rm c}, r_{\rm t}, f_{\rm cl},
\mu_{\rm sky})$? 2) How this accuracy depends on the photometric
signal to noise? 3) How degenerate are these parameters? 4) What is
the minimal signal to noise required to constrain the cluster's
half-light radius $r_{\rm h}$?

For simplicity, we do not take into consideration the stochastic
effects due to random sampling of stellar mass function. The
stochasticity of bright stars is significant for low mass star
clusters, therefore, the results of this study are primarily
applicable to massive clusters. The completely empty sky background
is realistic for the clusters in galaxy halos. However, even for
bright clusters projected on the galaxy we can assume that an
unresolved background stellar population is a uniform sky background
with elevated $\mu_{\rm sky}$ level.

\sectionb{2}{IMAGE SIMULATION}

We assume King (1962) surface brightness profile defined by its
central value, $\mu_0$, a core radius, $r_{\rm c}$, and a tidal
radius, $r_{\rm t}$:
\begin{equation}
\mu(r)=\mu_0\left[{\left({1+\frac{{r^2}}{{r_{\rm c}^2}}}\right)^{-1/2}-\left({1+\frac{{r_{\rm t}^2}}{{r_{\rm c}^2}}}\right)^{-1/2}}\right]^2.
\end{equation}

For further analysis we consider clusters with $r_{\rm c}$ = 3 pix
and $r_{\rm t}$ = 30 pix, which are extended objects with shallow
surface brightness profiles and affected by noise more than compact
clusters. Pixel values of a 2D array were computed using Eq.~1 by
placing a cluster in the center of the image, however, allowing its
position to be randomized within the pixel. The image is square with
a side length of $2\,r_{\rm t} + 1 = 61$ pix. It was convolved with
a Gaussian PSF of FWHM = 3 pixels. The total flux of the cluster was
set to $f_{\rm cl}$. A constant sky background, $\mu_{\rm sky}$, was
added and Poisson noise was introduced to each pixel to simulate
observations.

We define the photometric aperture of radius $r_{\rm phot}$ = 10
pix, which, for the chosen $r_{\rm c}$ = 3 and $r_{\rm t}$ = 30 pix
and the PSF, integrates 80\% of the cluster's total flux. We denote
this flux as $f_{\rm phot} = 0.8f_{\rm cl}$ and compute the signal
to noise ratio within the aperture following Newberry (1991):
\begin{equation}
S/N = \frac{f_{\rm phot}}{\sqrt{f_{\rm phot} + \pi \cdot r_{\rm phot}^2 \cdot \mu_{\rm sky}}}.
\end{equation}

The sky background level depends on the exposure time, telescope
size, passband, etc. We analyze three cases of the sky background
levels having flux ratios 1:10:100, which correspond to the the flux
ratios in galaxy halo, disk, and bulge, i.e., weak, moderate, and
strong. Therefore, the background levels were set to $\mu_{\rm sky}$
= $10^2$, $10^3$, and $10^4$ ADU (photons; inverse gain = 1
e$^-$/ADU).

We have chosen to investigate the cases $S/N$ = 5, 10, 20, 50, 100,
200, 500, and 1000 for each case of the sky background level by
adjusting the cluster's flux, $f_{\rm cl}$, to achieve the required
$S/N$. We have simulated the cluster grid with three sky and eight
signal to noise values; 24 nodes in total. For each node of the
grid, 100 cluster images were analyzed.

Examples of simulated clusters placed in sky background of $\mu_{\rm
sky}$ = $10^3$ ADU are displayed in Fig.~1 for eight cases of $S/N$.
The $S/N$ = 5 cluster is not detectable by eye, while that with
$S/N$ = 10 is barely distinguishable. Progressing to $S/N$ = 50, the
object is more prominent and suitable for model fitting. However,
the pixel values beyond the photometric aperture $r_{\rm phot}$ = 10
pix, shown as the inner circle, are dominated by noise even for the
$S/N$ = 200 cluster. When the $S/N$ = 1000 cluster is considered,
which is close to the maximum $S/N$ achievable in a single exposure
before the core of the cluster gets saturated, pixels close to the
tidal radius $r_{\rm t}$ = 30 pix, shown as the outer circle, are
still dominated by the noise. In this model (Eq.~1), pixels beyond
$r_{\rm t}$ contain photons from the sky background only.

\begin{figure}[!tH]\vbox{
\centerline{\psfig{figure=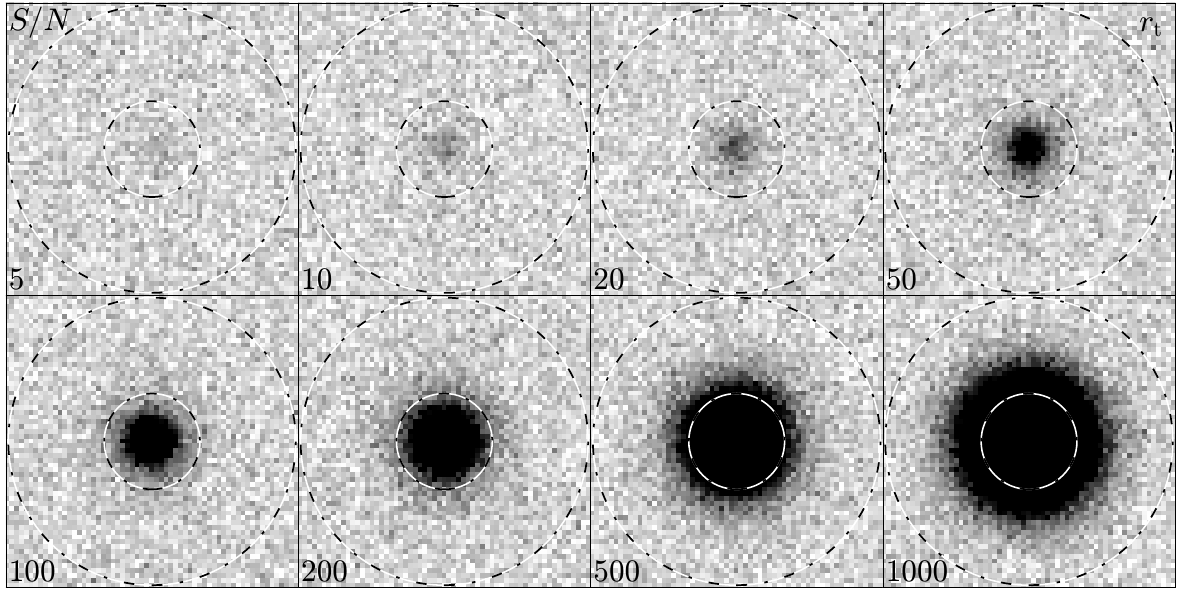,width=124mm}}
\captionb{1}{Examples of simulated star clusters placed in the sky
background $\mu_{\rm sky}$ = $10^3$ ADU. The signal to noise ($S/N$)
values (5, 10, 20, 50, 100, 200, 500, and 1000) are displayed in the
panels. Inner circle of $r_{\rm phot}$ = 10 pix indicates the
photometric aperture used to define the $S/N$ within it. Outer
circle indicates the tidal radius $r_{\rm t}$ = 30 pix ($r_{\rm c}$
= 3 pix). Grey levels correspond to pixel values from 950 to 1250
ADU in linear scale. The cluster's center with $S/N$ = 1000 has a
value of $\sim$41000 ADU.}}
\end{figure}

The other cases of the sky background levels, $\mu_{\rm sky}$ =
$10^2$ and $10^4$ ADU, show the results visually similar to those in
Fig.~1, however, when the sky level is high, it is a dominant source
of uncertainty even for brighter clusters.

\sectionb{3}{MODEL FITTING}

To perform star cluster analysis we use the Markov Chain Monte Carlo
(MCMC) model fitting technique implemented in the ``{\it emcee}''
algorithm (Foreman-Mackey et al. 2013). The model parameters are: 1)
position, $(x, y)$, 2) core and tidal radii, $r_{\rm c}$ and $r_{\rm
t}$, 3) cluster's total flux, $f_{\rm cl}$, and 4) sky background
level, $\mu_{\rm sky}$. To reduce computation time, we initialize
the ``{\it emcee}'' with true parameters of the simulated cluster.
The ``{\it emcee}'' starts a number of independent Markov chains
which sample the parameter space according to the goodness
criterion, which is the probability, $P$, that the observed image
can be explained by the model with given parameters.

We compute the probability of the model, $P$, by multiplying the
prior probability and the likelihood. We assume a flat prior
probability for the following parameter ranges: 1) position $(x, y)$
within 3 pixels of true position, 2) $0.1 < r_{\rm c} < r_{\rm t} <
200$ pix, 3) $10 < f_{\rm cl} < 10^9$ ADU, 4) $1 < \mu_{\rm sky} <
10^5$ ADU. If the Markov chain of ``{\it emcee}'' reaches the
boundary values, the prior probability 0 is returned and, therefore,
the probability of the model $P$ = 0. This allows the model to be
defined within reasonable limits of the parameter values, where the
probability of the model is equal to the likelihood of the model. We
build a smooth cluster model (Eq.~1) plus sky background as
described in the previous section to compute the likelihood.

To estimate model fitting uncertainties due to photon noise, we
consider the observed image and assume that for a pixel with flux
$\mu_{\rm obs}$ its uncertainty is $\sqrt{\mu_{\rm obs}}$ and is
equal to the standard deviation of the Gaussian distribution. This
assumption is reasonably applicable even for simulated images with
$\mu_{\rm sky}$ = $10^2$ ADU with Poisson noise. Computation of the
likelihood assuming a Gaussian distribution of uncertainty is much
faster than assuming a Poisson distribution.

We compute the Gaussian likelihood of the data pixel, given the
model pixel. We multiply the likelihoods of pixels assuming that
they are independent and assign the likelihood to the probability
$P$ because of a flat prior. The ``{\it emcee}'' uses $\ln(P)$ since
the numerical values of probability are small when dealing with a
large number of pixels. All pixels in the image are used for
computation. Pixels in the corners of the image are free of the
cluster's flux and allow to constrain the sky background as an
independent parameter.

We run the ``{\it emcee}'' for 3000 steps as a ``burn-in'' phase
while Markov chains reach the vicinity of the maximum likelihood
position in the parameter space. We drop these steps and run for
5000 steps more, which sample the posterior values of the model
parameters. These posterior distributions are analyzed in the
following section.

\sectionb{4}{RESULTS AND DISCUSSION}

We analyze the results of ``{\it emcee}'' model fitting by
producing corner plots, which show all the one- and two-dimensional
projections of the posterior probability distributions of the
parameters $(x, y, r_{\rm c}, r_{\rm t}, f_{\rm cl}, \mu_{\rm
sky})$, and demonstrate all the covariances between the parameters
and their marginalized distributions in histograms.

In Fig.~2 we show the posterior distributions of $r_{\rm c}$,
$r_{\rm t}$, $f_{\rm cl}$, and $\mu_{\rm sky}$ for the $S/N$ = 20
cluster displayed as black contours. True values are indicated by an
intersection point of large cross-bars. Open circles show median
values of the recovered parameter distributions of 100 simulated
clusters with the same input parameters, but different randomization
of Poisson noise. The radius $r_{\rm c}$ is recovered reasonably
well, however, judging by concentration of open circles at $r_{\rm
t} \sim 100$ pix (middle of $r_{\rm t}$ prior range), we conclude
that the $S/N=20$ is too low to constrain the tidal radius.
Non-linear degeneracies between the parameters, e.g., $r_{\rm c}$
and $r_{\rm t}$, are observed. The sky background is underestimated
and the cluster's flux is overestimated in the majority of cases.

\begin{figure}[!tH]\vbox{
\centerline{\psfig{figure=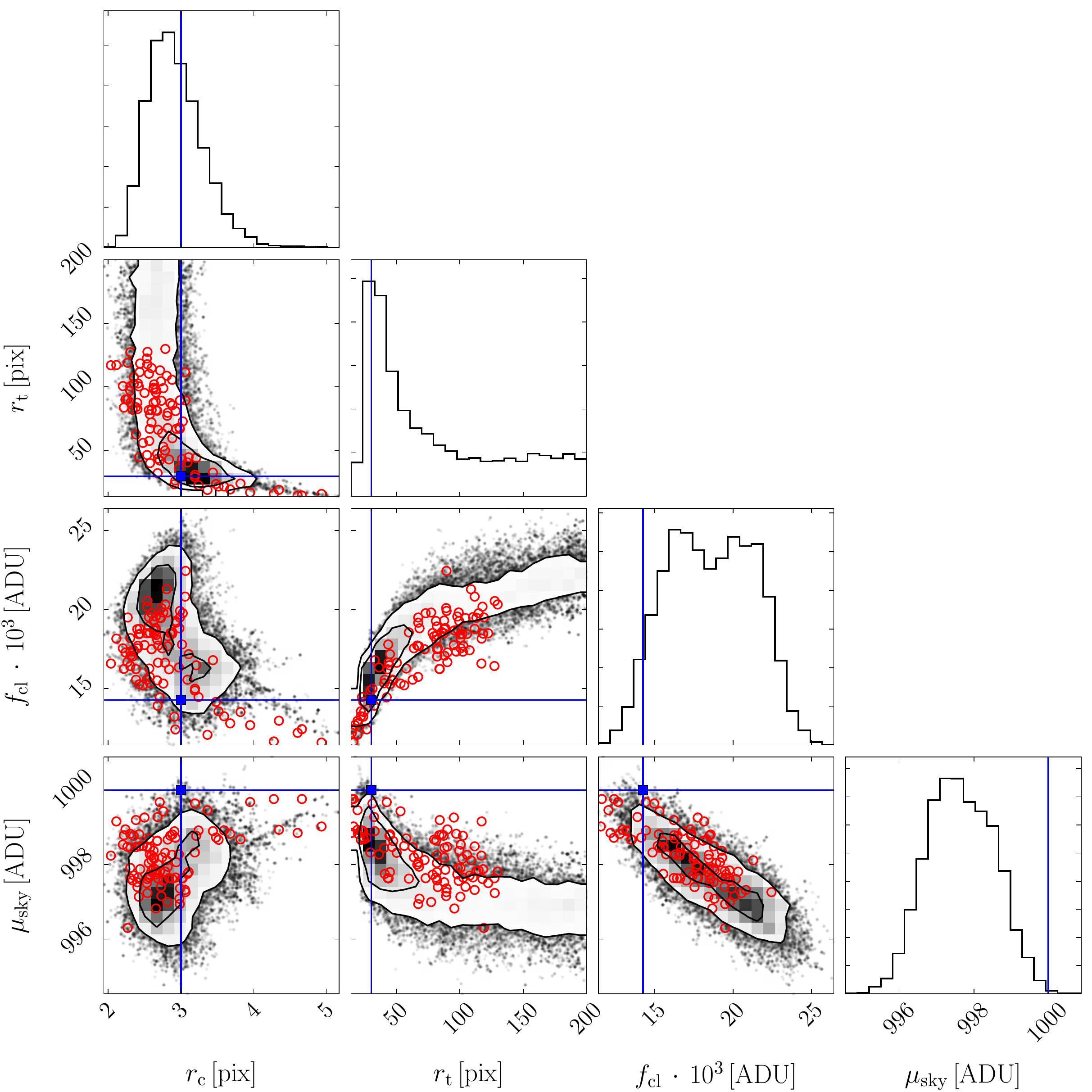,width=124mm}}
\captionb{2}{Posterior distributions of core, $r_{\rm c}$, and
tidal, $r_{\rm t}$, radii, flux $f_{\rm cl}$, and sky background
$\mu_{\rm sky}$, shown as histograms and two parameter covariance
plots for a star cluster with $S/N$ = 20 and $\mu_{\rm
sky}$\,=\,$10^3$ ADU. Contours are shown at 1- and 2-$\sigma$
levels. True parameter values are indicated by an intersection of
large cross-bars. Open circles show median values of the recovered
parameter distributions of 100 simulated clusters with the same
input parameters but different randomization of Poisson noise.}}
\end{figure}

When a cluster with larger $S/N$ is considered, see Fig.~3 for the
case $S/N$ = 50, the parameter degeneracies become almost linear and
histograms have less extended tails. The degeneracy between $r_{\rm
c}$ and $r_{\rm t}$ is prominent because models with small $r_{\rm
c}$ and large $r_{\rm t}$ are as likely as models with large $r_{\rm
c}$ and small $r_{\rm t}$. In the outskirts of the cluster, which
are dominated by the sky background noise, models with large $r_{\rm
t}$ are as likely as models with small $r_{\rm t}$, but there is
also a strong degeneracy between $r_{\rm t}$ and $f_{\rm cl}$,
because models with larger extent have a larger integral flux.

\begin{figure}[!tH]\vbox{
\centerline{\psfig{figure=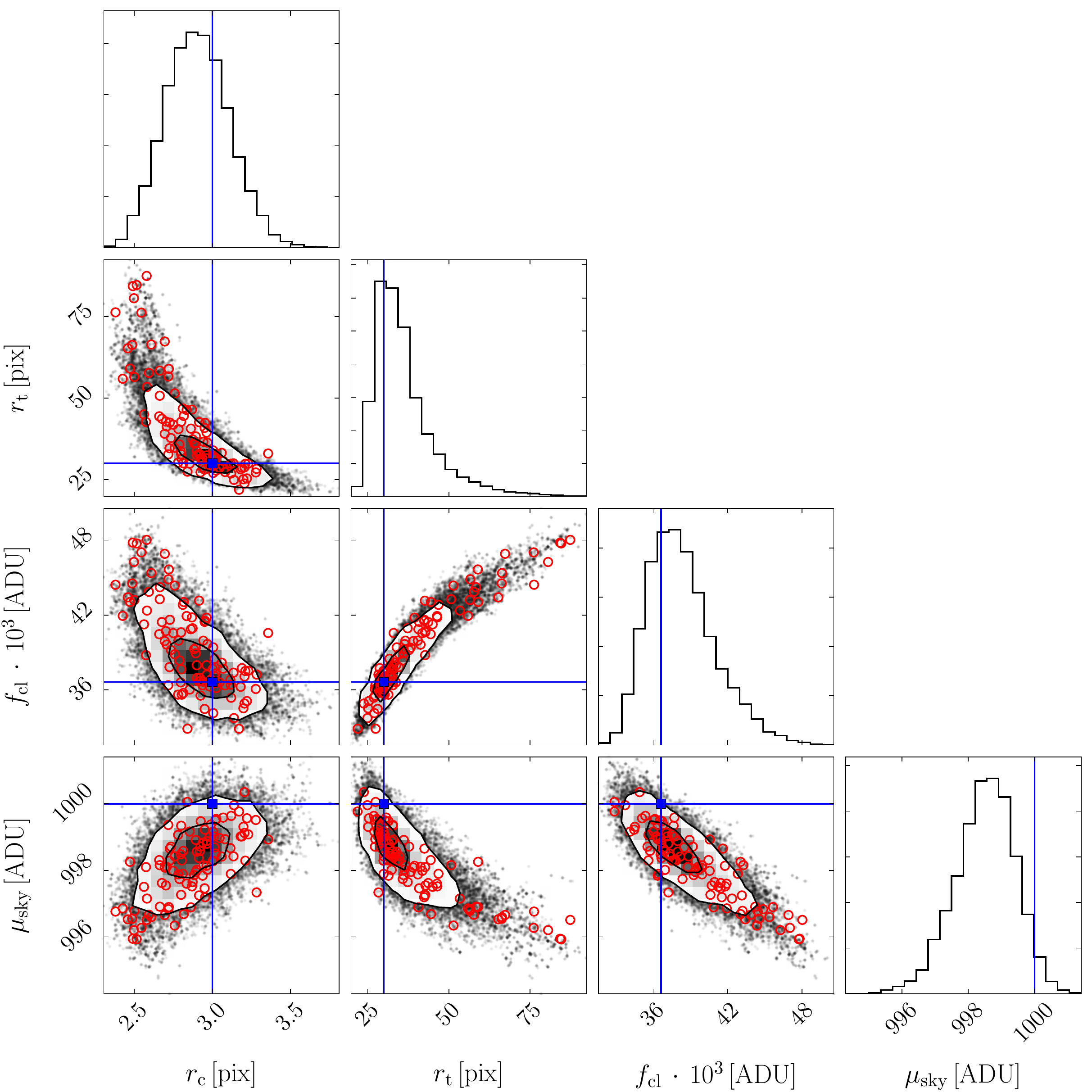,width=124mm}} \captionb{3}{Same
as in Fig. 2, but for $S/N$ = 50. Note the shorter range of axis
scales.}}
\end{figure}

Although beyond $r_{\rm t}$ photons only from the sky background are
detected, the majority of pixels in the image are within $r_{\rm
t}$, and the sky background is determined within the cluster's area.
Therefore, $\mu_{\rm sky}$ is also degenerate with $r_{\rm t}$,
because when a model with larger $r_{\rm t}$ is considered,
$\mu_{\rm sky}$ must be smaller then the true value in order to
compensate for the excess flux. The uncertainty in the sky
background remains identical to that in the $S/N$ = 20 case, because
it is determined primarily by the number of pixels that are
unaffected by the cluster's flux ($r > r_{\rm t}$).

For a cluster with $S/N$ = 100 (Fig.~4), the parameter degeneracies
are more pronounced, but parameter uncertainties are smaller.
Judging from the posterior distributions of the parameters of a
single cluster, displayed as contours in Figs. 2--4, and the
distributions of median values of posteriors of 100 clusters with
the same input parameters (open circles), we see a good agreement,
suggesting that the assumption of Gaussian uncertainty of pixel
values used in model fitting is valid. The uncertainty in the
derived parameter values can in this case be judged from the ``{\it
emcee}'' posterior samples of a single cluster.

\begin{figure}[!tH]\vbox{
\centerline{\psfig{figure=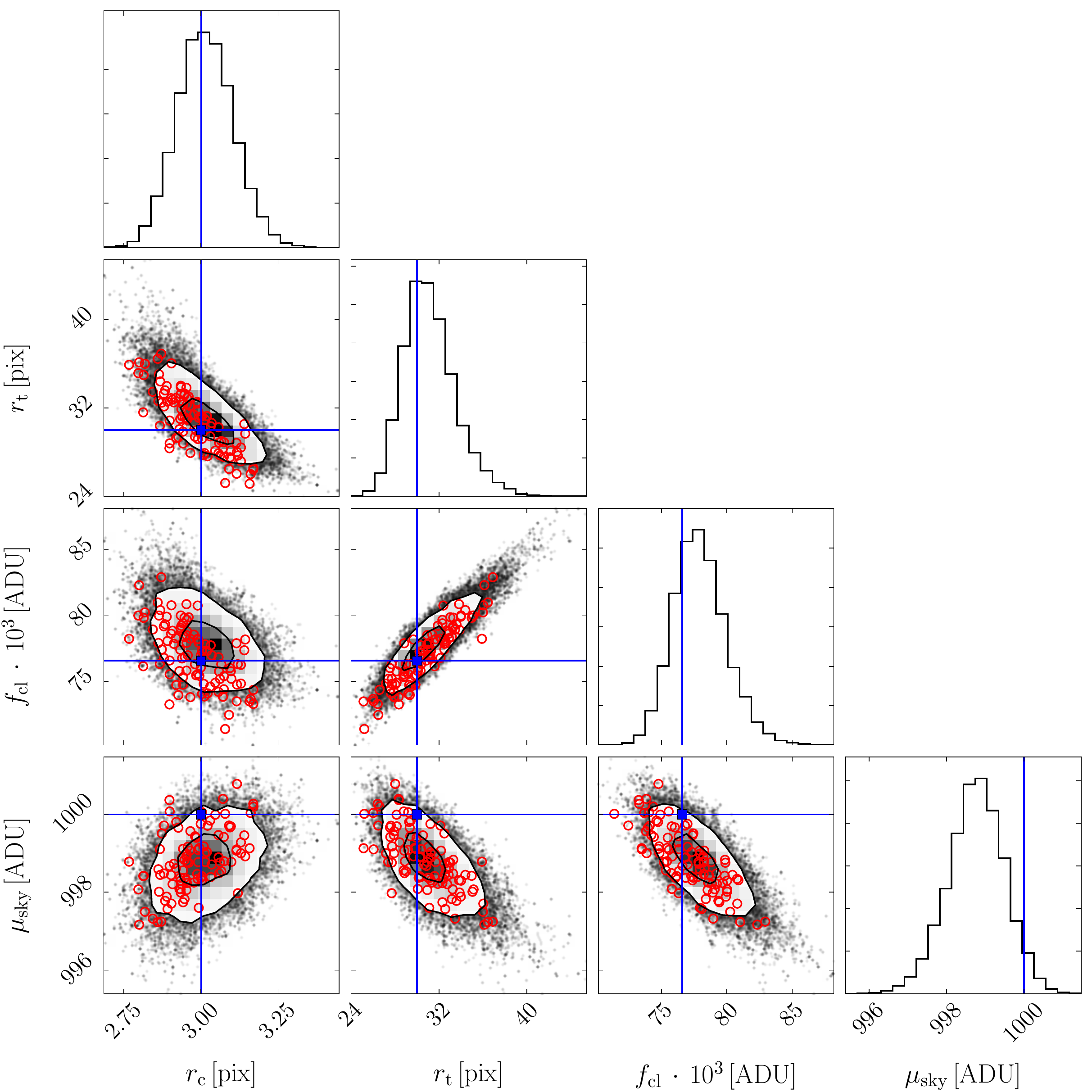,width=124mm}} \captionb{4}{Same
as in Fig. 2, but for $S/N$ = 100. Note the shorter range of axis
scales.}}
\end{figure}

We introduce the half-light radius, $r_{\rm h}$, in the analysis,
since in numerous studies it is reported as cluster's size
indicator. By analytically integrating Eq.~1, we get an expression
for $r_{\rm h}$ as a function of $(r_{\rm c}, r_{\rm t})$. This
expression is solved numerically to yield $r_{\rm h}$. Once the
posterior values of $r_{\rm c}$ and $r_{\rm t}$ are obtained by the
``{\it emcee}'', the posterior values of $r_{\rm h}$ are computed.

We compute median values of the parameter ($r_{\rm c}, r_{\rm t},
r_{\rm h}$) posterior distributions for each node cluster from the
model grid with sky background values of $\mu_{\rm sky}$ = $10^2$,
$10^3$, and $10^4$ ADU and $S/N$ = 20, 50, 100, 200, and 500 (100
clusters per node). Then we compute the median and the standard
deviation of each parameter per node (same $\mu_{\rm sky}$ and
$S/N$) and plot these values as a function of $S/N$ in Fig.~5. We
see that for $S/N$ = 20 clusters, $r_{\rm t}$ is overestimated and
$r_{\rm c}$ is underestimated because of a large degeneracy of these
parameters as seen in Fig.~2. To constrain the cluster's half-light
radius to be accurate to within $\sim$20\%, $S/N\sim50$ is necessary
in all realistic sky background level conditions ($10^2$, $10^3$,
and $10^4$ ADU).

\begin{figure}[!tH]\vbox{
\centerline{\psfig{figure=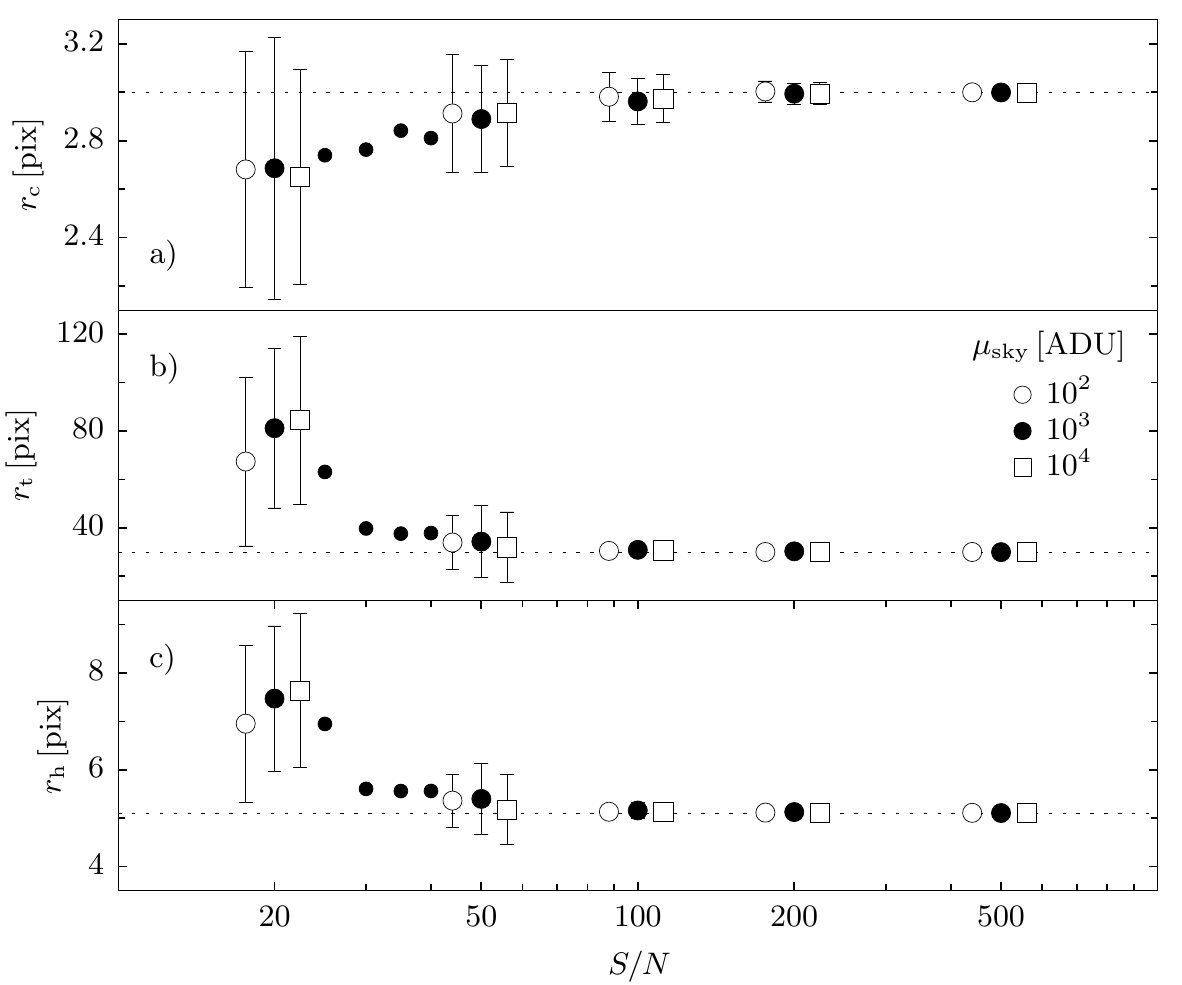,width=120mm}} \captionb{5}{Median
and standard deviation of the recovered radii $r_{\rm c}$, $r_{\rm
t}$, and $r_{\rm h}$ of 100 clusters per model grid node as a
function of signal to noise, $S/N$. For the same $S/N$, symbols
indicate different sky background, $\mu_{\rm sky}$ = $10^2$ (open
circles), $10^3$ (large filled circles), and $10^4$ ADU (squares).
Horizontal dashed lines show true parameter values. Additional
simulations at intermediate values of $S/N$ = 25, 30, 35, and 40
with $\mu_{\rm sky}$ = $10^3$ ADU are shown by small filled
circles.}}
\end{figure}

Bonatto \& Bica (2008) presented the results of their study on the
accuracy of structural parameters of star clusters recovered from
the radial profiles of light, mass, and star counts and their
dependence on photometric depth. They assumed that sky background
level is known, subtracted it from the data and performed analysis
of radial profiles only.

Carlson \& Holtzman (2001) performed an extensive analysis of the
possibility of measuring King (1962) model parameters as a function
of cluster's $r_{\rm c}$, $r_{\rm t}$, and signal to noise. They
used a Levenberg-Marquard optimization algorithm, which provides the
best-fit value of the model parameters. Only three free parameters
were used to fit the model: position ($x, y$), and $r_{\rm c}$
($r_{\rm c}/r_{\rm t}$ was fixed). In this paper, we demonstrate the
posterior distributions of all parameters that were used to generate
the cluster's model, allowing also the sky background as a free
parameter. By comparing the results for 100 clusters with the same
parameters, we demonstrate that posterior distributions of a single
cluster are a good indicator of true parameter uncertainty, if the
image pixel uncertainty distribution is Gaussian.

\sectionb{5}{CONCLUSIONS}

Using the MCMC model fitting technique we have analyzed the
influence of photometric signal to noise on the accuracy of derived
structural parameters of unresolved star clusters. We have
considered clusters with a King (1962) model surface brightness
profile ($r_{\rm c}$ = 3 pix, $r_{\rm t}$ = 30 pix) and convolved it
with a Gaussian PSF of FWHM = 3 pix. We have used the ``{\it
emcee}'' program to perform MCMC model fitting for the simulated
grid of artificial clusters with various values of sky background
and signal to noise.

We have demonstrated the degeneracies between the structural
parameters, cluster's flux, and sky background level. By defining
the photometric aperture containing 80\% of the cluster's flux, we
have found that $S/N\sim50$ is critical to constrain tidal radius if
the cluster resides in sky background of $10^3$ ADU. We have found
that $S/N\sim50$ is necessary in all realistic sky background level
conditions ($10^2$, $10^3$, and $10^4$ ADU) to constrain the
cluster's half-light radius to an accuracy better than $\sim$20\%.
The presented technique can be applied to synthetic images
simulating various observations of extragalactic star clusters.

\thanks{We thank the referee for comments and suggestions which helped
to improve the manuscript. This research was funded by a grant No.
MIP-102/2011 from the Research Council of Lithuania.}

\References
\refb Bonatto C., Bica E. 2008, A\&A, 477, 829
\refb Carlson M. N., Holtzman J. A. 2001, PASP, 113, 1522
\refb Foreman-Mackey D., Hogg D. W., Lang D., Goodman J. 2013, PASP, 125, 306
\refb King I. 1962, AJ, 67, 471
\refb Newberry M. V. 1991, PASP, 103, 122
\end{document}